\documentclass[twocolumn,showpacs,preprintnumbers,amsmath,amssymb]{revtex4}

\usepackage{graphicx}
\usepackage{dcolumn}
\usepackage{bm}

\begin{document}

\title{State transition of a non-Ohmic damping system\\ in a corrugated plane}
\author{Kun L\"u and Jing-Dong Bao}
\thanks{Corresponding author. Electronic mail: jdbao@bnu.edu.cn}
\affiliation {Department of Physics, Beijing Normal University,
Beijing 100875, China}
\date{\today}

\begin{abstract}
Anomalous transport of a particle  subjected to non-Ohmic damping of
the power $\delta$ in a tilted periodic potential is investigated
via Monte Carlo simulation of generalized Langevin equation. It is
found that the system exhibits two relative motion modes: the
locking state and the running state. Under the surrounding of
sub-Ohmic damping ($0<\delta<1$), the particle should transfer into
a running state from a locking state only when local minima of the
potential vanish; hence the particle occurs a synchronization
oscillation in its mean displacement and mean square displacement
(MSD). In particular, the two motion modes are allowed to coexist
 in the case of super-Ohmic damping
($1<\delta<2$) for moderate driving forces, namely, where exists
double centers in the velocity distribution. This induces the
particle having faster diffusion, i.e., its MSD reads $\langle\Delta
x^2(t)\rangle=2D^{(\delta)}_{\textmd{eff}}t^{\delta_{\textmd{eff}}}$.
Our result shows that the effective power index
$\delta_{\textmd{eff}}$ can be enhanced and is a nonmonotonic
function of the temperature and the driving force. The mixture
effect of the two motion modes also leads to a breakdown of
hysteresis loop of the mobility.

\end{abstract}

\maketitle

\section{INTRODUCTION}

There are many physical situations which can be described by
Brownian transport in a tilted periodic potential, for example,
  Josephson junction \cite{Bar-Book}, charge-density wave
\cite{Gru-Prl}, superionic conductor \cite{Ful-Prl}, rotation of
dipoles in external field \cite{Reg-Pre}, phase-locking loop
\cite{Lin-Book}, diffusion on surface \cite{Fre-Prl}, and separation
of particles by electrophoresis \cite{Adj-Book}. The quantitative
properties of those systems have been discussed in plenty of
literatures \cite{Ris-Book}, such as the dependence of the coherence
level of transport on the temperature, driving force and shape of
potential \cite{Hei-arXiv}; the huge enhancement of the effective
diffusion coefficient relative to the free diffusion
\cite{Rei-Prl,Dan-Pre,Rei-pre}; the response
 of output to the noise and signal
\cite{Reg-Europhys.Lett.}, and so on \cite{Sas-arXiv}. Since
theoretical tools and numerical algorithms are not sufficient  in
non-Markovian dynamics with a frequency-dependent non-Ohmic damping,
most of the models are established in the Ohmic damping environment.
However, the frequency-dependent  damping is more general because a
large number of stochastic processes fail to be the Markovian
dynamics.

Recent studies on anomalous diffusion and transport are mostly
limited in the absence of potential, linear force case or the
sub-diffusion in a potential \cite{Hei-Pre}. It is worth to point
out that the behavior of a particle moving in a periodic potential
immersed in the super-Ohmic damping environment might be far more
complicated than that in the Ohmic and sub-Ohmic damping cases.
 In comparison with the previous findings for great enhancement of
the effective diffusion coefficient \cite{Rei-Prl,Dan-Pre,Rei-pre}
and the hysteresis loop of mobility \cite{Ris-Book} in the Ohmic
damping environment, we will perform a detailed investigation in the
present work on the diffusion and the mobility of a particle
subjected to arbitrary non-Ohmic damping in a corrugate plane. This
is in terms of
 an effective algorithm proposed by us \cite{Lv-Bao} to
numerically solve a generalized Langevin equation (GLE) with
arbitrary damping kernel function and corresponding thermal colored
noise.

The paper is organized as follows. In Sec. II, we describe briefly
the anomalous transport model by means of the GLE. In Sec. III, the
two basic quantities of interest: the generalized effective
diffusion coefficient and the fractional mobility are defined; the
novel behaviors of diffusion and mobility are shown. Finally, we
draw a conclusion of our findings in the section IV.

\section{THE MODEL}

We consider a Brownian particle moving in a one-dimensional periodic
potential under the influence of non-Ohmic memory friction and a
constant external driving force. The dynamics of the particle is
governed by the following GLE \cite{Kubo-book1,Kubo-book2},
\begin{eqnarray}
\dot x(t)&=&v(t),\\
\nonumber
m\dot{v}(t)&=&-m\int_{0}^{t}\gamma(t-t')v(t')dt'+U'(x)+\sqrt{m
k_BT}\xi(t),
\end{eqnarray}
where $k_B$ is the Boltzmann constant, $T$ is the temperature of the
environment, $\gamma(t)$ is the damping kernel function and related
to $\xi(t)$ through the fluctuation-dissipation theorem
\cite{Kubo-book1,Mur-Pra}
\begin{equation}
\langle\xi(t)\xi(t')\rangle=\gamma(|t-t'|),
\end{equation}
where $\xi(t)$ is a zero mean Gaussian colored noise and its
spectral density reads
\begin{equation}
\langle|\xi(\omega)|^2\rangle=2\gamma_\delta\left(\frac{|\omega|}{\tilde{\omega}}\right)
^{\delta-1}f_c\left(\frac{|\omega|}{\omega_c}\right).
\end{equation}
The small $|\omega|$ behavior of $\langle|\xi(\omega)^2|\rangle$ is
a power-law characterized by the index $\delta-1$. The function
$f_c(|\omega|/\omega_c)$ is a high frequency cutoff function of
typical width $\omega_c$ \cite{Gra-Prl}, and
$\tilde{\omega}\ll\omega_c$ denotes a reference frequency allowing
for the constant $m\gamma_\delta$ to have a dimension of viscosity
for any $\delta$ \cite{Wei-book}. The cases of $0<\delta<1$ and
$1<\delta<2$ are  the sub-Ohmic damping and super-Ohmic damping,
respectively; $\delta=1$ is the Ohmic one, then $\gamma(\omega)$ is
equal to a constant and the noise is white. In Eq. (1), $U(x)$ is
considered to be a tilted periodic potential,
\begin{equation}
U(x)=-U_0\cos\left(\frac{2\pi}{\lambda} x\right)-Fx.
\end{equation}
The minima of $U(x)$ vanish when the driving force $F$ is taken to
be the critical value: $F_c=U_02\pi/\lambda=1.0$.

In the calculation, the  natural unit ($m=1$ and $k_B=1$), the
dimensionless parameters: $U_0=1.0$, $\lambda=2\pi$,
$\gamma_\delta=4.0$,  the smooth cutoff function
$f_c=\exp(-\omega/\omega_c)$ \cite{Wei-book} with $\omega_c=4.0$,
and the time step $\Delta t=0.01$ are used.  The test particles
start from the origin of coordinate and have zero velocity, here
$2\times 10^4$ test particles are used to describe the stochastic
distribution of a Brownian particle.

\section{\label{sec:level3}diffusion and Mobility}

The quantities of foremost interest are the diffusion coefficient
and the mobility. Here we generalize the both into non-Ohmic damping
case with an arbitrary power index $\delta$,
\begin{eqnarray}
&&D^{(\delta)}:=\lim_{t\rightarrow
\infty}\frac{1}{\Gamma(1+\delta)}{_0D}_t^\delta\langle\Delta
x^2(t)\rangle_\delta,\\  &&\mu_{\delta}:=\lim_{t\rightarrow
\infty}\frac{1}{F\sin(\frac{\delta\pi}{2})}{_0D}_t^\delta \langle
x(t)\rangle_\delta,
\end{eqnarray}
where ${_0D}_t^\delta$ denotes the fractional derivative. The
algorithm for numerically calculating the two quantities is
presented in the Appendix \ref{sec:level6}.

\subsection{\label{sec:level42}Diffusion}

\begin{figure}
\includegraphics[width=0.48\textwidth,height=0.4\textwidth]{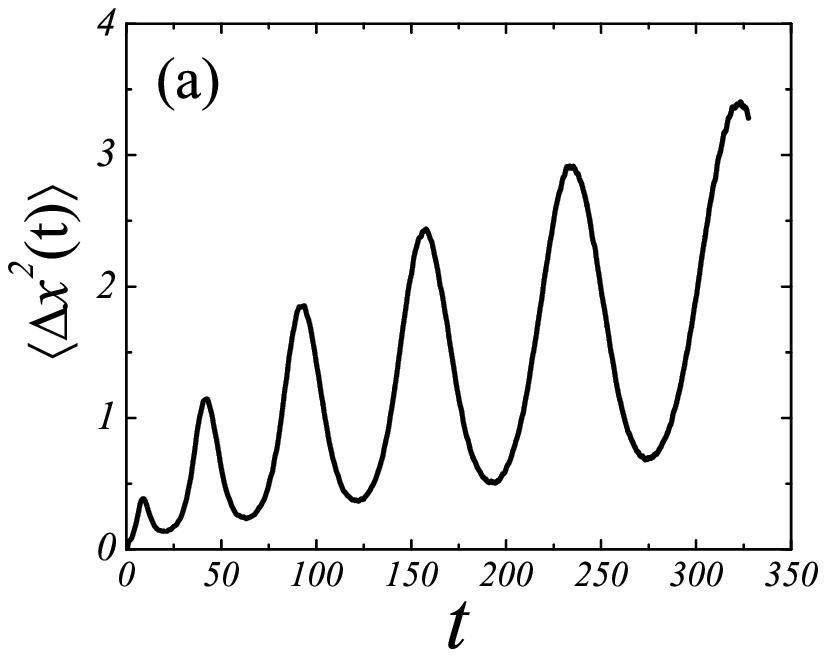}
\includegraphics[width=0.48\textwidth,height=0.4\textwidth]{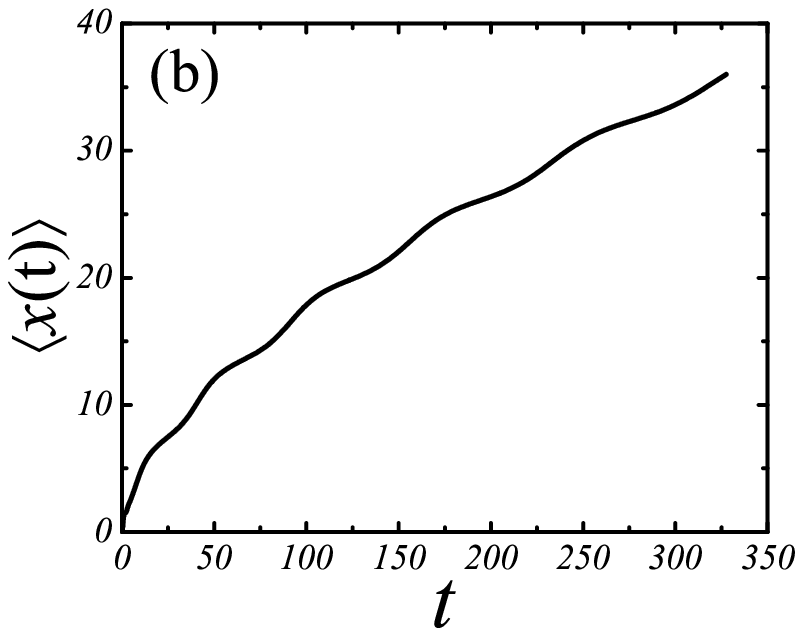}
\caption{(a) Time-dependent MSD of the sub-Ohmic damping particle of
$\delta=0.6$ and (b) its mean displacement. The parameters used are:
 $F=5.0$ and $T=0.1$.}
\end{figure}

We begin our studies from the situation of sub-Ohmic damping. We
have noticed that the case of sub-diffusion dynamics has been
discussed by Goychuk and H\"{a}nggi \cite{Goychuk-arXiv}, where the
GLE and the fractional Fokker-Planck equation approaches to the
escape dynamics are used and compared. The escape is governed
asymptotically by a power law whose exponent depends exponentially
on both, the barrier height and the temperature. If the ratio of the
barrier height to temperature is too large, the diffusion motion in
a washboard potential well below a critical tilt cannot be observed
numerically within a reasonable time window, i.e., nearly all the
test particles are confined in the locking state. Therefore, we
consider the transport of sub-Ohmic damping particle subjected to a
large driving force because of the efficiency of numerical
simulation of GLE. Figures 1 (a) and 1 (b) show time-dependent MSD
and MD of the particle of $\delta=0.6$. With increasing the driving
force $F$ until local minima of the potential vanish, we find that
the MSD of the particle shows a quasi-periodic oscillation. As long
as the
 MSD of the particle experiences a quasi-period process,
the particle will move the distance of a periodic length $\Delta
x=2\pi$ along the direction of external force.

\begin{figure}
\includegraphics[width=0.48\textwidth,height=0.4\textwidth]{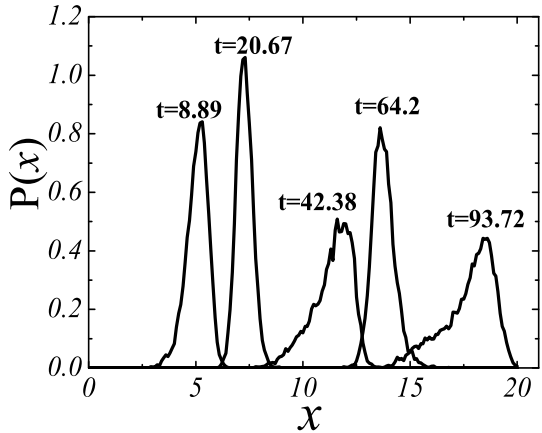}
\caption{ The space probability distribution of the sub-Ohmic
damping particle of $\delta=0.6$ at different times for $F=5.0$ and
$T=0.1$.}
\end{figure}

In Fig. 2, we plot the space probability distribution of a sub-Ohmic
damping particle at different times. It is seen that with the
evolution of time, the width of the probability distribution becomes
periodically narrow when the particle moves in the bottom of a
potential well; the one is broad when the particle arrives at the
top of potential. Unlike the normal Ohmic damping particle
\cite{Rei-Prl}, its probability distribution is not centralized.
 Our results can be interpreted as follows. Under the sub-Ohmic
damping environment, the particle has a strong memory to its initial
position and thus the diffusion in the coordinate space is much
slow. If the potential have local minima, it is quite difficult for
the particle to escape from the potential well, thus the particle is
in a locking state during the period of simulation. As the local
minima of the potential vanish, the particle subjected to a large
driving force can enter the running state. Therefore, its
distribution width is modulated by the periodic structure of
potential and thus the MSD of particle occurs a quasi-periodic
oscillation.

\begin{figure}
\includegraphics[width=0.48\textwidth,height=0.4\textwidth]{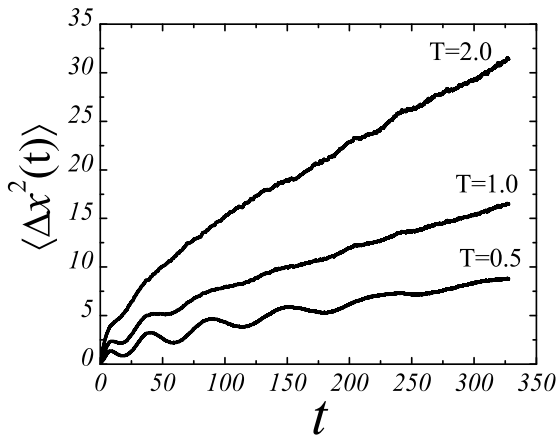}
\caption{ The MSD of the particle with $\delta=1.7$
 at a large driving force $F=5.0$ for various temperatures.}
\end{figure}

 Figure 3 shows that the
quasi-periodic oscillation phenomenon becomes unconspicuous when the
temperature increases for $T>1.0$ at $F=5.0$.  In this case the
particle transfers completely into the running state  and hence the
structure of the potential might have less influence on the
transport process.

\begin{figure}
\includegraphics[width=0.48\textwidth,height=0.4\textwidth]{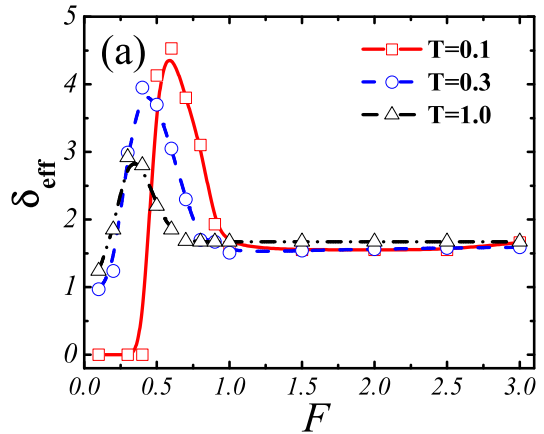}
\includegraphics[width=0.48\textwidth,height=0.4\textwidth]{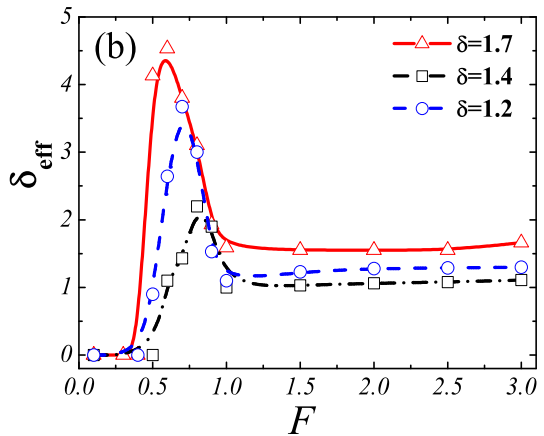}
\caption{(Color online) (a) The effective power index
$\delta_{\textmd{eff}}$ vs $F$ for various $T$ at $\delta=1.7$.  (b)
The effective power index vs $F$ for various $\delta$ at $T=0.1$.}
\end{figure}

In particular, in the case of super-Ohmic damping, we find
numerically that the asymptotic MSD of the particle can be
approximately written as a power function,
\begin{equation}
\langle \Delta x^2(t)\rangle=2D^{(\delta)}_{{\textmd{eff}}}
t^{\delta_{\textmd{eff}}(T,F)},
\end{equation}
 where $\delta_{\textmd{eff}}$ depends on $T$ and
$F$. Indeed, the index $\delta_{\textmd{eff}}$ is not always equal
to $\delta$ like the Ohmic and the sub-Ohmic damping cases, but
varies non-monotonically with $F$. For a moderate $F$, we find a
prominent result: The effective power  index $\delta_{\textmd{eff}}$
exceeds $2$ (i.e., the ballistic diffusion
\cite{Bao-Prl,Bao-Pre1,Bao-Pre2}) when the periodic potential is
titled observably but its local minima still exist. Further analysis
shows that the mysterious diffusion behavior is caused by the mixing
of the locking state and the running state.

In Figs. 4 (a) and 4 (b), we plot the effective power index
$\delta_{\textmd{eff}}$ as a function of $F$ for various $T$ and
$\delta$. It is seen from Fig. 4 (a)
 that the maximal value of $\delta_{\textmd{eff}}$ versus $F$
forward with increasing temperature. Similar behavior of
$\delta_{\textmd{eff}}$ for other super-Ohmic damping cases can also
be observed, as shown in Fig. 4 (b). The smaller the value of
$\delta$ is, the larger $F$ where the maximum
$\delta_{\textmd{eff}}$ appears. This can be interpreted
qualitatively as follows. In the non-Markovian rate theory
\cite{Hanggi-RMP}, for a sufficiently big ratio of barrier height to
temperature, the super-diffusion in a periodic potential should turn
out into the normal diffusion because the escape events are
exponentially distributed in time and no overlong jumps can occur,
so that $\delta_{\textmd{eff}}=1$ when the tilt of the potential is
small. Also, for very large $F$ and the structure effect of
potential vanishing, all the test particles can move into a running
state, thus $\delta_{\textmd{eff}}=\delta$.
 However, for a middle tilt, some test particles
are confined in the potential well (in the locking state) and others
drift quickly forward (in the running state).  A proper proportion
between the locking state and the running state should induce the
maximum of the effective  power index. Apparently it becomes easier
for the particle in the locking state to escape the well and join
the running state with the increase of either $T$ or $\delta$.
Therefore, the maximum of $\delta_{\textmd{eff}}$ appears in the
case of a small $F$ at high temperature; in the case of large
$\delta$ at low temperature, respectively.

\begin{figure}
\includegraphics[width=0.48\textwidth,height=0.4\textwidth]{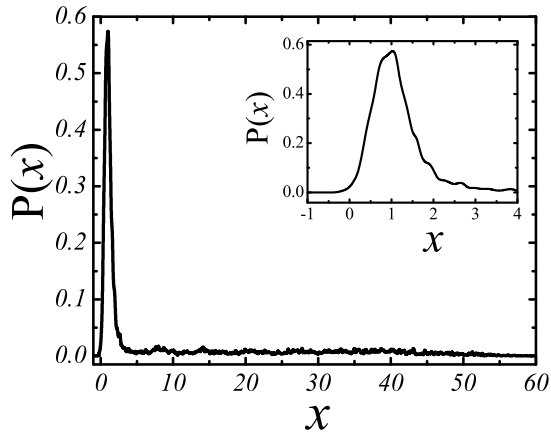}
\caption{The space probability distribution of the particle  at time
$t=50.0$. The inset figure is the probability distribution at the
locking state only. The parameters used are: $\delta=1.7$, $F=0.75$
and $T=0.1$. }
\end{figure}

\begin{figure}
\includegraphics[width=0.48\textwidth,height=0.4\textwidth]{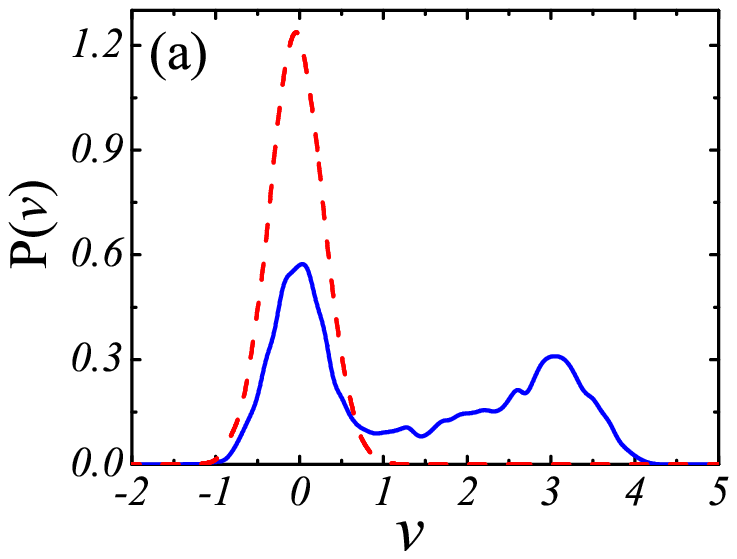}
\includegraphics[width=0.48\textwidth,height=0.4\textwidth]{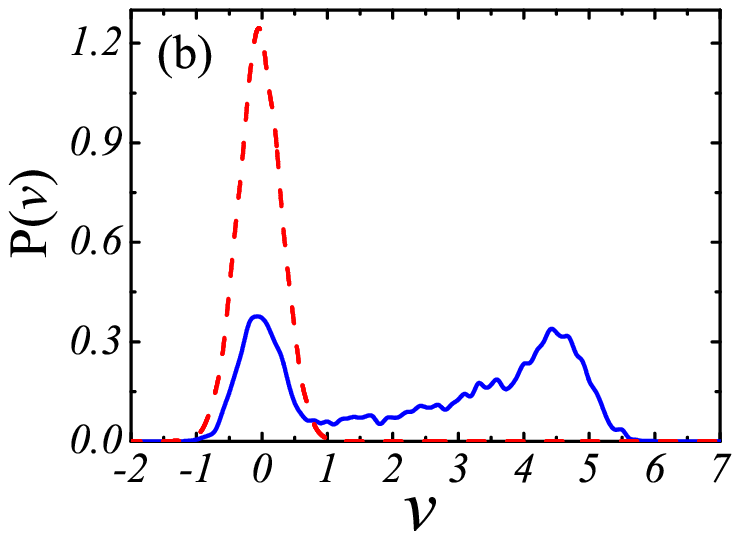}
\caption{(Color online) The velocity distributions of the particle
at $t=150$ (a) and $t=330$ (b). The solid and dashed lines are the
results of the supper-Ohmic $\delta=1.7$ and Ohmic $\delta=1.0$
cases, respectively. The parameters used are:  $F=0.75$ and
$T=0.1$.}
\end{figure}

In Fig. 5, we illustrate the case of $\delta=1.7$ at the low
temperature ($T=0.1$) and the middle tilt ($F=0.75$) to depict the
coexistence of the two motion modes.  The backward and forward
barrier heights are given by
\begin{eqnarray}
U_1&=&2U_0\sqrt{1-\left(\frac{F\lambda}{2\pi
U_0}\right)^2}+\frac{F\lambda}{\pi}
\arcsin\left(\frac{F\lambda}{2\pi U_0}\right)
+\frac{F\lambda}{2}, \nonumber\\
U_2&=&2U_0\sqrt{1-\left(\frac{F\lambda}{2\pi
U_0}\right)^2}+\frac{F\lambda}{\pi}\arcsin\left(\frac{F\lambda}{2\pi
U_0}\right) -\frac{F\lambda}{2}.
\end{eqnarray}
For a low temperature $T<U_2\ll U_1$, the particle oscillates around
the potential well, however, it still escapes over a barrier with a
small probability. Since the barrier crossing process  is quite slow
at low temperature, the particle in the locking state has an
approximate Gaussian space distribution
 centering at $x_0$
[$x_0=\frac{\lambda}{2\pi}\arcsin(\frac{F\lambda}{2\pi U_0})$], as
shown in the inset of Fig. 5. Once the test particle climbs over the
barrier, it will no longer be restricted again. Because the next
hill of the titled periodic potential is lower than the present one,
the kinetic energy of the particle gaining from the external driving
force is greater than the dissipated energy due to the memory
friction. The particle can enter the running state, so it drifts
quickly along the direction of driving force. This results in a long
tail appearing in the space probability distribution of the
particle. For the Ohmic damping particle, once escaping over a
barrier, it will slide to the next well and be trapped again. Hence
the space probability distribution of the Ohmic damping particle is
smashed into some small pieces and does not make up a running state.

Figures 6 (a) and 6 (b) show the coexistence of the two motion modes
of the super-Ohmic damping particle, which can be more intuitively
for the velocity distributions of the particle at times $t=150$ and
$t=330$. As expected, we do not find the coexistence phenomenon of
two velocity modes appearing in the normal diffusion. It is seen
that the super-Ohmic damping particle with an increasing probability
enters the running state; the difference between the two center
velocities in the running and rocking states becomes large with the
evolution of time.  This implies that as long as the test particles
escape out of the well, they will be accelerated by the corrugated
plane and join the running state.  Of course, the present locking
state occurs simply because we cannot observe the motion on the
time-scale of our numerical simulation as the escape rate are very
rare \cite{Goychuk-arXiv}, but a new locking state relative to the
forward running state should arise once the original locking state
disappears.

\begin{figure}
\includegraphics[width=0.48\textwidth,height=0.4\textwidth]{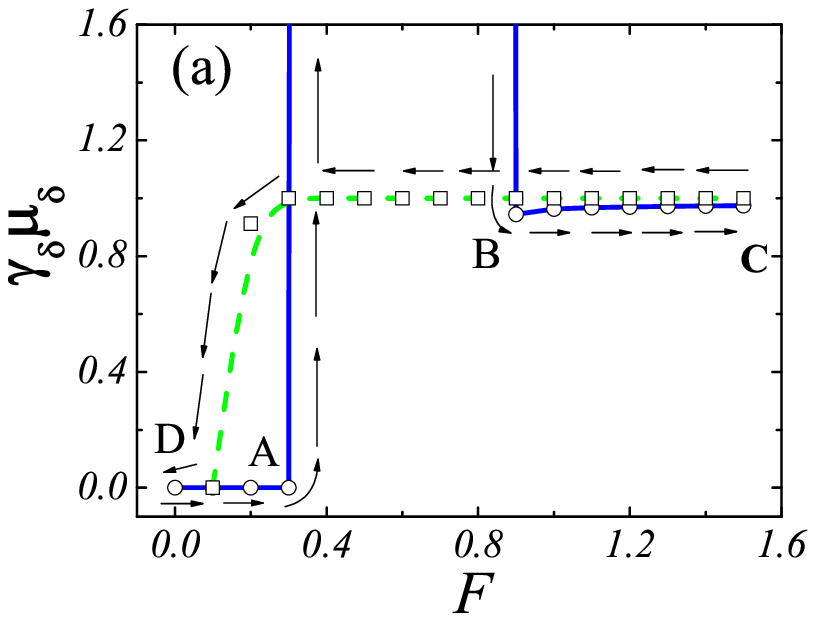}
\includegraphics[width=0.48\textwidth,height=0.4\textwidth]{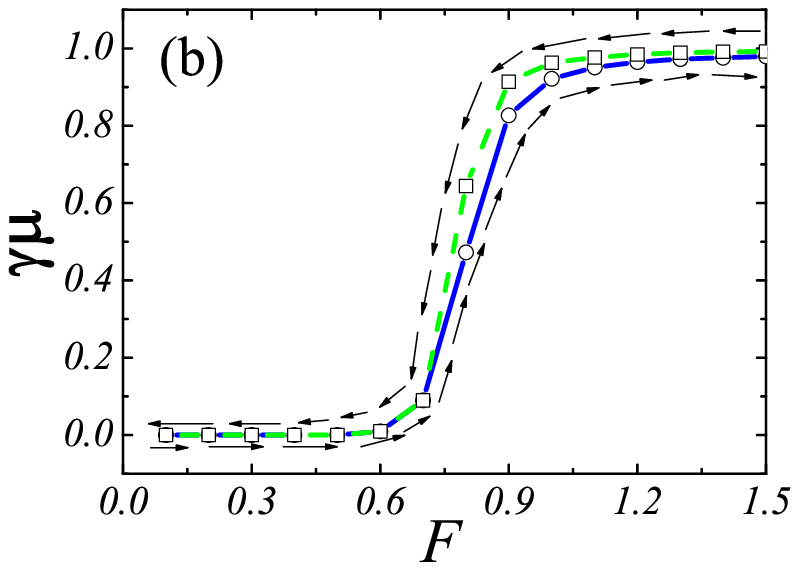}
\caption{(Color online) (a) The fractional mobility times the
damping constant as a function of driving force in the super-Ohmic
damping case with $\delta=1.7$. (b) The normal Ohmic damping result
with $\gamma=0.5$. The solid and dashed lines correspond to the
forward and backward processes, respectively. The temperature is
$T=0.1$.}
\end{figure}

\subsection{Mobility}

The mobility determined by Eq. (6) as a function of the driving
force is plotted in Fig. 7. Starting from zero tilt and switch
adiabatically on the tilt $F$, the mobility of the particle remains
zero when all the test particles are in the locking state until some
of them join the running state.  In comparison with the Ohmic
damping case, we find that the hysteresis loop is broken and become
stagger. At that point $A$, the mobility jumps to infinity and then
drops to a constant at point $B$ and keeps this constant with
increasing $F$ (point $C$). The point $B$ corresponds to the
critical force $F=F_c$, where the local minima of the corrugated
plane vanish. When the driving force decreases adiabatically, all
the test particles are kept in the running state and the mobility
approaches a constant until the driving force becomes so small that
most of the test particles are trapped in the potential wells. At
the point $D$, the mobility falls to zero again. While for the usual
Ohmic damping case shown in Fig. 7 (b), we find that a bistability
occurs in the region $0.7<F<1.4$, which forms a hysteresis loop.
However, for the sub-Ohmic damping case, we have not found similar
hysteresis loop of mobility as what arises in the Ohmic damping
case.

\section{\label{sec:level5}summary}

We have investigated the transport of a non-Ohmic damping  particle
in a titled periodic potential and reported a prominent finding: The
titled periodic potential as a simple equipment which not only
enhances the diffusion coefficient, but also changes diffusive
behavior of the particle. This is due to a novel phenomenon of two
motion modes: the locking state and the running state,  which can
appear and transfer in the corrugated plane. In the sub-Ohmic
damping case, the mean square displacement of the particle shows a
quasi-periodic property
 when the driving force is larger than the critical value where the minima
 of the potential
 vanish.
 While for the super-Ohmic damping case, the two
motion modes, namely, there exists two centers in the velocity
distribution, can be coexisted and transferred. Thus the power index
for the mean square displacement of the particle is enhanced. In
comparison with the hysteresis loop of mobility of the normal case,
the hysteresis loop of mobility of a super-Ohmic damping particle is
broken.

The anomalous Brownian motion in a periodic potential is
representable for many applications occurring in areas such as in
condensed matter physics, chemical physics, molecular biology,
communication theory, and so on. We are confident that both
theoretical and experimental works in the future will help one to
clarify further and shed more light onto all these intriguing issues
and problems.

\section * {ACKNOWLEDGEMENTS}
 This work was supported by the National
Natural Science Foundation of China under Grant No. 1067401 and the
Specialized Research Foundation for the Doctoral Program of Higher
Education under Grant No. 20050027001.

\appendix{}
\section{\label{sec:level6}Numerical methods for fractional
calculus}

The so-called Riemann-Liouvill fractional integral is defined
through \cite{Sam-Book}
\begin{equation}
_{t_0}I_t^{-\delta}f(t)=\frac{1}{\Gamma(\delta)}\int_{t_0}^tdt'\frac{f(t')}{(t-t')^{1-\delta}},t>0,\delta>0,
\end{equation}
whereas its left-inverse $_{t_0}D_t^\delta$ reads
\begin{equation}
_{t_0}D_t^\delta:=_0D_t^m{_{t_0}I_t}^{\delta-m},m-1<\delta<m,m\in\mathbb{N},
\end{equation}
where $_0D_t^m$ denotes the ordinary derivative of order $m$. In
this present work, the case of $t_0=0$ is concerned. For
completeness, we define
\begin{equation}
_0I_t^0=_0D_t^0=\textbf{I},
\end{equation}
where $\textbf{I}$ is the identity operator. It is convenient to
make use of the discrete operators of translation (shift) and finite
differences to derive the approximative recursive expressions to the
fractional differentiation operator $_0D_t^{\delta}$. In a lucid way
the theory of numerical differentiation and integration (with
equidistant grid points) has been developed in Chapters 7 to 10 of
Ref. \cite{C.E.-Book}. See also Chapter 6 of Ref. \cite{Isa-Book}.

Let $\tau\in \mathbb{R}$, we define the shifting operator $E^\tau$
and the backward difference operator $\nabla_\tau$ by them acting on
a function $u(t)$ for $t\in \mathbb{R}$,
\begin{eqnarray}
&&E^\tau u(t)=u(t+\tau),\\
\nonumber &&\nabla_\tau u(t)=u(t)-u(t-\tau).
\end{eqnarray}
We furthermore have the relation, with $\textbf{I}$ as the identity
operator,
\begin{equation}
\nabla_\tau=\textbf{I}-E^{-\tau}.
\end{equation}
Using these notations, we write the  approximation
$[u(t)-u(t-\tau)]/h$ for the derivative $u'(t)$ of a differentiable
function $u(t)$ as $\nabla_h u(t)/h$ for a small positive $h$ with
accuracy $\textit{O}(h)$ as the function $u(t)$ is sufficiently
smooth. High order derivatives $u^{(n)}(t)=_0D_t^n u(t)$ (
$n\in\mathbb{N}$) with small $h>0$, can be approximated by
\begin{equation}
[\nabla_h^{(n)}u(t)]/h^n=h^{-n}(\textbf{I}-E^{-h})^nu(t),
\end{equation}
again in case of $u(t)$ being sufficiently smooth, with order of
accuracy $\textit{O}(h)$. The powers $\nabla_h^{(n)}$ can be
readily expanded via the binomial theorem
\begin{equation}
\nabla_h^{(n)}=\sum_{j=0}^n(-1)^j\bigg({n \atop j}\bigg)E^{-jh}.
\end{equation}
This leads to the known formula
\begin{equation}
h^{-n}\sum_{j=0}^n(-1)^j\bigg({n \atop
j}\bigg)u(t-jh)=_0D_t^nu(t)+\textit{O}(h).
\end{equation}

The remarkable fact now is that these formulas can be generalized to
the case of non-integer order of derivative. Replacing the positive
integer $n$ by a positive real number $\delta$ amounts to use the
formal power
\begin{equation}
\nabla_h^\delta=\sum_{j=0}^\infty(-1)^j\bigg({\delta \atop
j}\bigg)E^{-jh},
\end{equation}
in analogy to the expansion ($E^{-h}$ replaced by the complex
variable $z$)
\begin{equation}
(1-z)^\delta=\sum_{j=0}^\infty(-1)^j\bigg({\delta \atop
j}\bigg)z^j,
\end{equation}
which is convergent if $|z|<1$. We thus get the
G\"{r}unwald-Letnikov approximation:
\begin{equation}
h^{-\delta}\nabla_h^\delta
u(t)=h^{-\delta}\sum_{j=0}^\infty(-1)^j\bigg({\delta \atop
j}\bigg)u(t-jh)=_0D_t^\delta u(t)+\textit{O}(h).
\end{equation}
If $u(t)$ decays to zero sufficiently fast as $t\rightarrow \infty$
, in particular if $u(t)=0$ for $t<0$, $\nabla_h^\delta$ will not
diverge and hence for the latter case, we have
\begin{equation}
h^{-\delta}\nabla_h^\delta
u(t)=h^{-\delta}\sum_{j=0}^{[t/h]}(-1)^j\bigg({\delta \atop
j}\bigg)u(t-jh).
\end{equation}

By using the property of Gamma function
\begin{displaymath}
\Gamma(\delta)\Gamma(1-\delta)=\frac{\pi}{\sin(\pi\delta)},
\end{displaymath}
we obtain the final recursion of calculating fractional calculus:
\begin{equation}
_0D_t^\delta
u(kh)=\frac{h^{-\delta}}{\Gamma{(-\delta)}}\sum_{j=0}^{k-1}\frac{\Gamma(j-\delta)}{\Gamma(j+1)}u((k-j)h).
\end{equation}

\end{document}